%Paper: hep-th/9305055
%From: QUEVEDO <Fernando.Quevedo@iph.unine.ch>
%Date: Thu, 13 May 1993 11:14:37 +0200

%%% this paper has two figures appended in a second part.
%\def\figin#1{\vskip2in} % puts blank space instead of figures
%%% if you do not  have epsf.tex, then comment  the following line:
\input epsf \def\figin#1{#1}

\input harvmac

\def\s{\sigma}
\def\L{\Lambda}

\def\d{{\rm d}}
\def\IR{\relax{\rm I\kern-.18em R}}
%\font\cmss=cmss10 \font\cmsss=cmss10 at 7pt
\def\IZ{\relax\ifmmode\mathchoice
{\hbox{\cmss Z\kern-.4em Z}}{\hbox{\cmss Z\kern-.4em Z}}
{\lower.9pt\hbox{\cmsss Z\kern-.4em Z}}
{\lower1.2pt\hbox{\cmsss Z\kern-.4em Z}}\else{\cmss Z\kern-.4em Z}\fi}

\def\delbar{\overline\del}
\def\inbar{\vrule height1.5ex width.4pt depth0pt}
\def\IC{\relax\thinspace\hbox{$\inbar\kern-.3em{\rm C}$}}
\def\figinsert#1#2#3{\topinsert\figin{#2}\centerline{\vbox{\baselineskip12pt
\advance\hsize by -1truein\noindent\footnotefont{\bf Fig.~\xfig#1:} #3}}
\bigskip\endinsert}

%\font\largerb = cmss10 scaled 1700

\Title{\vbox{\baselineskip12pt\hbox{NEIP93-003}\hbox{hepth/9305055}}}
{\vbox{\centerline{Abelian and Non--Abelian}
   \vskip2pt\centerline{Dualities in String Backgrounds   }}}\bigskip

\centerline{ Fernando Quevedo\footnote{*} { Talk presented  at the
 Workshop {\it From Superstrings to Supergravity}, Erice 1992.
 Supported by Swiss National Science Foundation.
  E--mail: quevedo@iph.unine.ch }}\smallskip
\bigskip
\centerline{\it Institut de Physique}\smallskip
\centerline{\it Universit\'e de Neuch\^atel}\smallskip
%\baselineskip=12pt
%\centerline{\it Universit\'e de Neuch\^atel}
%\centerline{Rue A.-L. Breguet 1}
%\baselineskip=12pt
\centerline{\it CH-2000 Neuch\^atel, Switzerland}
\bigskip\medskip\bigskip\bigskip\medskip
%\vglue 0.8cm
%\centerline{\tenrm ABSTRACT}
%\vglue 0.3cm
%{\rightskip=3pc
% \leftskip=3pc
% \tenrm\baselineskip=12pt%\parindent=1pc
 \noindent
We present a brief discussion of recent work on
 duality symmetries in non--trivial string backgrounds.
Duality is obtained from a gauged non--linear $\s$--model
with vanishing gauge field strength. Standard results are
reproduced for abelian gauge groups, whereas a new type of
duality is identified for non--abelian gauge groups. Examples of duals
of WZW models and 4--d black holes are given.
 %\vglue 0.8cm
%}

\Date{5/93}

\noblackbox

\lref\bn{I. Bars and D. Nemechansky, Nucl. Phys. B348 (1991) 89.}
\lref\witten{E. Witten, Phys. Rev. {\bf D44} (1991) 314.}
\lref\rwzw{E. Witten, Comm. Math. Phys. {\bf 92} (1984) 455.\semi
For a review see P. Goddard and D. Olive,
Int. J. Mod. Phys. {\bf 1} (1986) 303.}
\lref\wzwg{E. Witten, Nucl. Phys. {\bf B223} (1983) 422\semi
K. Bardacki, E. Rabinovici and B. Saering,
Nucl. Phys. {\bf B301} (1988) 151 \semi
D. Karabali and H. J. Schnitzer,
Nucl. Phys. {\bf  B329} (1990) 649, and references therein.}
\lref\busc{T. Buscher, Phys. Lett. {\bf 194B} (1987) 59;
 Phys. Lett. {\bf 201B} (1988) 466.}
\lref\vhol{For a review see, J. W. van Holten, Z. Phys. {\bf C27:57} (1985).}
\lref\gmv{M Gasperini, J. Maharana and G. Veneziano,
Phys. Lett. {\bf 272B} (1991) 277.}
\lref\sen{A. Sen, Phys. Lett. {\bf 271B} (1991) 295; Phys. Lett.
 {\bf 274B}(1992) 34
\semi
S. Hassan and A. Sen, Nucl. Phys. {\bf  B375} (1992) 103.}
\lref\md{M. Duff, talk at Trieste summer 1989,
Nucl. Phys. {\bf  B335} (1990) 610.}
\lref\muller{M. Muller, Nucl. Phys. B337 (1990) 37.}
\lref\cfmp{C. Callan, D. Friedan, E. Martinec and M. Perry,  Nucl.
Phys. {\bf B262} (1985) 593.}
\lref\kir{E.B. Kiritsis, Mod. Phys. Lett. {\bf A6} (1991) 2871.}
\lref\vero{M. Ro\v cek and E. Verlinde, Nucl. Phys. {\bf B373 } (1992)
 630.}
\lref\giva{P. Ginsparg and C. Vafa, Nucl. Phys. {\bf B289} (1987) 414 \semi
    T. Banks, M. Dine, H. Dykstra and
    W. Fischler, Phys. Lett. {\bf 212B} (1988) 45\semi
    E. Alvarez and M. Osorio, Phys. Rev. {\bf D40} (1989) 1150.}
\lref\grvsw{A. Giveon, E. Rabinovici and G. Veneziano,
Nucl. Phys. {\bf B322} (1989) 167\semi
A. Shapere and F. Wilczek, Nucl. Phys. {\bf B320} (1989) 669.}
\lref\gz{M.K. Gaillard and B. Zumino, Nucl. Phys. {\bf B193} (1981) 221.}
\lref\cfg{S. Cecotti, S. Ferrara and Girardello,
Nucl. Phys. {\bf B308} (1988) 436.}
\lref\iltq{L.E. Ib\'a\~nez, D. L\"ust, F. Quevedo and S. Theisen
unpublished (1990) \semi G. Veneziano, Phys. Lett. B265 (1991) 287.}
\lref\bmq{C. Burgess, R. Myers and F. Quevedo unpublished (1991) \semi
A. Tseytlin, Mod. Phys. Lett. A6 (1991) 1721.}
\lref\bv{R. Brandenberger and C. Vafa, Nucl. Phys. B316 (1989) 391.}
\lref\mv{K. Meissner and G. Veneziano, Phys. Lett. {\bf B267} (1991) 33\semi
M. Gasperini and G. Veneziano, Phys. Lett. {\bf B277} (1992) 256.}
\lref\tv{A. Tseytlin and C. Vafa, Harvard preprint HUTP-91/A049
(hepth@xxx/\-9109048).}
\lref\dvv{R. Dijgkraaf, E. Verlinde, and H. Verlinde,
Nucl. Phys. {\bf B371} (1992) 269\semi
A. Giveon, Mod. Phys. Lett. {\bf A6} (1991) 2843.}
\lref\rAGG{L. Alvarez-Gaum\'e and P. Ginsparg,
Ann. Phys. 161 (1985) 423.}
\lref\rpglh{P. Ginsparg, ``Applied conformal field theory,''
Les Houches lectures (summer, 1988), published in
{\it Fields, Strings, and Critical Phenomena\/},
ed.\ by E. Br\'ezin and J. Zinn-Justin, North Holland (1989).}
\lref\rvafa{C. Vafa, ``Topological Mirrors and Quantum Rings'',
Harvard preprint HUTP-91/A059 (hepth@xxx/\-9111017).}
\lref\px{P. Candelas, X.C. de la Ossa, P.S. Green and
L. Parkes, Nucl. Phys. B359 (1991) 21; Phys. Lett. 258B (1991) 118.}
\lref\barst{I. Bars,
University of Southern California preprint USC-91/HEP-B4.}
\lref\hhs{J.H. Horne and G. Horowitz,
Nucl. Phys.  B368 (1992) 444
\semi J. Horne, G. Horowitz and A. Steif,
Phys. Rev. Lett. 68 (1992) 568.}
\lref\hhss{J. Horne, G. Horowitz and A. Steif,
University of California preprint
UCSBTH-91-53 (hepth@xxx/9110065).}
\lref\giro{A. Giveon and M. Ro\v cek, Nucl. Phys. {\bf B380} (1992) 123.}
\lref\ind {S.P. Khastgir and A. Kumar Bubaneswar, preprint
(hepth@xxx/9109026).}
\lref\cresc{M. Crescimanno, Berkeley preprint LBL-30947\semi
P. Ho\v rava Chicago preprint EFI-91-57 (hepth@xxx/9110067)\semi
I. Bars and K. Sfetsos, Univ. Southern California preprints
USC-91/HEP-B5 (hepth@xxx/\-9110054) and USC-91/HEP-B6
(hepth@xxx/\-9111040)
\semi D. Gershon, preprint TAUP-1937-91 (hepth@xxx/9202005).}
\lref\dlp{L. Dixon, J. Lykken and M. Peskin, Nucl. Phys.
B325 (1989) 325.}
\lref\barsc{I. Bars, Nucl. Phys. B334 (1990) 125.}
\lref\hel{S. Helgason, ``Differential Geometry, Lie Groups,
and Symmetric Spaces'', Academic Press (1978)\semi
R. Gilmore, ``Lie Groups, Lie Algebras and Some of Their
Applications'', Wiley (1974).}
\lref\host{G. Horowitz and A. Strominger,
Nucl. Phys. B360 (1991) 197.}
\lref\gique{P. Ginsparg and F. Quevedo,  Nucl. Phys. {\bf B 385} (1992) 527,
 and
references cited therein.}
\lref\einstein{C. Hoenselaers and W. Dietz (eds.) ``Solutions
of Einstein's Equations: Techniques and Results '', Springer Verlag,
Berlin (1984) .}
\lref\nicolai{H. Nicolai, ``Two--Dimensional Gravities and
Supergravities as Integrable Systems'', preprint DESY 91-038 (1991).}
\lref\narain{K. S. Narain, Phys. Lett. Bxxx, (1986)}
\lref\flst{S. Ferrara, D. L\"ust, A. Shapere and S. Theisen, Phys. Lett.
B225 (1989) 363\semi E. J. Chun, J. Mas, J. Lauer and H.P. Nilles,
Phys. Lett. B233 (1989) 141\semi M. Cveti\v c, A. Font, L.E. Ib\'a\~nez,
D. L\"ust and F. Quevedo, Nucl. Phys. B361 (1991) 194.}
\lref\bigs{S. Weinberg, Field Theory Notes (1984).}
\lref\bigsd{See for instance, S. Weinberg, ``Gravitation and Cosmology'',
Wiley (1972).}
\lref\bars{I. Bars and K. Sfetsos,
Phys. Lett. {\bf 277B} (1992) 269; Mod. Phys. Lett. {\bf A7} (1992) 109; USC
preprints USC-92-HEP-B1, B2.}
\lref\gibbons{G.W. Gibbons and K. Maeda,
Nucl. Phys. {\bf B298} (1988) 741\semi
  D. Garfinkle, G. Horowitz and A. Strominger,
  Phys. Rev. {\bf D43} (1991) 3140.}
\lref\duality{K. Kikkawa and M. Yamasaki,
Phys. Lett. {\bf 149B} (1984) 357\semi
N. Sakai and I. Senda, Progr. Theor. Phys. Suppl. {\bf 75} (1986) 692.}
\lref\tseyt{A. Tseytlin, ``Cosmological Solutions with Dilaton and
Maximally Symmetric Space in String Theory'', Cambridge preprint
DAMTP--15--1992 (1992). }
\lref\nappi{C. R. Nappi, Phys. Rev. D21 (1980) 418\semi
     B. E. Fridling and A. Jevicki, Phys. Lett. 134B (1983) 70\semi
     E. S. Fradkin and A. A. Tseytlin, Ann. Phys. 162 (1985) 48.}
\lref\tseytd{A. A. Tseytlin, Mod. Phys. Lett. A6 (1991) 1721\semi
            A. S. Schwarz and A. A. Tseytlin, preprint
	    Imperial/TP/92-93/01.}
\lref\dlp{L. Dixon, J. Lykken and M. Peskin, Nucl. Phys. {\bf B325} (1989)
325.}
\lref\barsc{I. Bars, Nucl. Phys. {\bf B334} (1990) 125.}
\lref\deque{X.C. de la Ossa and F. Quevedo, Neuch\^atel preprint
NEIP92--004
(1992). Nucl. Phys. {\bf B} to appear, and references therein.}
\lref\bsn{I. Bars and Sfetsos ....}
\lref\bsnn{I. Bars and Sfetsos.......}
\lref\tseytn{A. Tseytlin .....}
\lref\hh{J.H. Horne and G. Horowitz,
Nucl. Phys. {\bf B368} (1992) 444.
}
\lref\kt{C. Kounnas and A. Tseytlin, contributions to these proceedings.}
\lref\hs{C. Hull and B. Spence, Phys. Lett. {\bf 232B} (1989) 204
\semi
        I. Jack, D. Jones, N. Mohammedi and H. Osborn, Nucl. Phys.
 {\bf B332} (1990) 359. }

\newsec{Introduction}

String Theory and Physics are at the moment two unrelated subjects
which are expected to have an interesting connection.
The only available approach towards this connection
 is to study the structure and physical properties of
the semiclassical string vacua given by classes of conformal field
theories
(CFT). For applications to Particle Physics, it is enough to assume
a flat four-dimensional spacetime and parametrize the different models
by an
internal CFT restricted by some consistency conditions such as
worldsheet modular invariance. In this way many classes of models have
been
studied and some, close to the Standard Model of Particle Physics give
hope that strings may actually be related to the real world.
Furthermore, duality and mirror symmetries have been identified
in these vacua and play an important role in determining the
low energy properties of the theory.  To ask questions
about gravitation in the context of String Theory, we have to relax the
flat spacetime assumption and substitute it by a general noncompact
CFT. These questions are of fundamental importance since the main
motivation for studying string theories is to provide a consistent way of
quantizing gravity. In particular it is of prime importance to
study singularities of cosmology and black hole-type  of
geometries in the context of String Theory, since it is  in those regimes
that the standard field theory methods of General Relativity fail to apply.
Duality, being a property of all string vacua with abelian
isometries \busc, has also been found in these backgrounds and
  could provide  a way  to understand how string
theory probes those singular geometries. A generalization of this
duality symmetry
to backgrounds with non--abelian isometries was recently discovered
 \deque. In this talk we will briefly discuss these
developments.

\newsec{Non--Compact String Backgrounds}

We know several ways  to construct non--compact
string vacua. The straightforward approach is to look for
solutions of the string background equations. Except for a few
exceptions \kt,
these are solutions of only
the
 lowest order in $\alpha '$  equations  \cfmp\ :
\eqn\einst{R_{MN} + D_M D_N \Phi -
            {1\over 4}H_M^{LP} H_{NLP}=0}
\eqn\hmn{D_L H^L_{MN}-(D_L\Phi)H^L_{MN}=0}
\eqn\dil{R-2\Lambda-(D\Phi )^2 +
            2D_M D^M\Phi-{1\over 12}H_{MNP}H^{MNP}=0\ ,}
where $\L\equiv (c-26)/3$ is the cosmological constant in the
effective string action, $c$ is the central charge and, as usual
$H_{MNP}\equiv \del_{[M} B_{NP]}$.  For the heterotic
string, these equations
are modified in order to include the background gauge fields.
{}From this we can extract the trivial but  powerful
conclusion that all
 solutions of Einstein's equations in vacuum
 are also solutions of the leading order string
background equations with constant dilaton and antisymmetric
tensor field.
Therefore we already have a large class of  solutions
to these equations \einstein. In particular the  4--d
Schwarzschild black hole geometry is a solution to the leading
order string equations,
when tensored with an appropriate CFT to provide the correct
 central charge.
This is not the case however for the Maxwell--Einstein system which
is generally modified by the dilaton coupling to the gauge fields
and then the Reissner--Nordstrom geometry, for instance, does not lead
to
a string background. The corresponding stringy solution
is the so--called charged dilatonic black hole of reference \gibbons.

The second approach is to look for non--compact CFT's directly.
In this case non--compact cosets are the appropriate class
of models to study. However, contrary to the compactification
approach, we cannot only construct  CFT's and use their algebraic
properties, in the non--compact case we have to provide a geometrical
interpretation to those CFT's.

In order to have a geometrical interpretation of these CFT's, we need to
have a Lagrangian formulation of them, which is naturally provided by
the WZW construction.
The WZW action for a group $G$ with
elements $g(z,\bar z)$ in complex coordinates is \rwzw,
\eqn\wzcc{L(g)={k\over{4\pi}}\int \d^2z\, \tr(g\inv\del g\, g\inv\delbar g)-
{k\over{12\pi}}\int_{B}\tr (g\inv dg\wedge g\inv dg\wedge g\inv dg)\ ,}
where the boundary of $B$ is the 2D worldsheet.
It is known that this action provides the conserved currents $g\inv\del g$
and $\delbar g g\inv$ to satisfy a chiral algebra  and then
giving a CFT  for the group $G$.
For the coset $G/H$ \wzwg\ the standard way in nonlinear sigma models
\vhol\ is to eliminate
the $H$ degrees of freedom by gauging the
 subgroup $H$ as a symmetry of \wzcc.
To promote the global $g\to h_L\inv\, g\, h_R$ invariance to a local
$g\to h\inv_L(z)\,g\,h_R(\overline z)$ invariance, we let
$\del g\to \del g + Ag$, and $\delbar g\to \delbar g-g\bar A$. The gauge
 fields
transform as $A\to h_L\inv(A+\del)h_L$ and $\bar A\to h_R\inv(\bar
A+\delbar)h_R$ (so that $Dg\to h_L\inv Dg\, h_R$ for $D$ equal to either
holomorphic or anti-holomorphic covariant derivative).
Vector gauge transformations correspond to $h_L=h_R$, and axial gauge
transformations to $h_L=h_R\inv$.
For abelian groups $H$,  both vector and axial-vector gauging are
anomaly free. In the non-abelian case,
only the vector gauging is allowed.
The gauged action may
be written as
\eqn\wzwcc{L(g,A)=L(g)+{k\over{2\pi}}\int \d^2z\,
 \tr\bigl( A\,\delbar g g\inv\mp\bar A \,g\inv
 \del g +A \bar A \mp g\inv A g \bar A \bigr)\ ,}
where the upper and lower signs represent respectively vector ($g\to hgh\inv$)
and axial-vector ($g\to hgh$) gauging.

\def\tdbhtext{The causal structure of the two dimensional black hole spacetime
of \witten. Regions I,IV are asymptotic regions, regions II,III are inside
the horizon, and regions V,VI are beyond the singularities.}

We now consider some naive properties of the geometry described by \wzwcc\
in the large $k$  limit.
Writing $A=A^a\sigma_a$ in terms of the generators $\sigma_a$ of $H$,
and integrating out the components $A^a$ classically gives the effective
action
\eqn\effcc{L=L(g)\pm{k\over{2\pi}}\int \d^2z\
 \tr(\s_b g\inv\del g)\, \tr(\s_a\delbar g g\inv)
 \,\Lambda_{ab}\inv\ ,}
with $\Lambda_{ab} \equiv \tr(\s_a \s_b \mp \s_a g \s_b g\inv)$.
This action can be identified with a $\sigma$--model action of the form
\eqn\efft{S=\int \d^2z\,\bigl(G_{MN}+B_{MN}\bigr)\,\del X^M \delbar X^N}
to read off the background metric and antisymmetric tensor field (torsion).

 Notice that
singularities of $\Lambda$ occur at least at fixed points of the gauge
transformation $g\to h g h^{\mp1}$. This is because for infinitesimal
$h\approx 1+\alpha^a\,\sigma_a$, we see that a fixed point $g$ satisfies
$\sigma_a\,g\mp g \sigma_a=0$. Multiplying by $g\inv \sigma_b$ and taking
the
trace, we see that $\Lambda=0$ at a fixed point.

It can be seen that,
 in the case of $H$ abelian the ungauged axial or vector symmetry
remains a {\it global\/} symmetry, i.e.\ an isometry of the spacetime
geometry. In the non-abelian case, not even a global vestige of the
ungauged symmetry remains (unless $H$ commutes with a subgroup of $G$).
In the abelian case, this implies that a fixed
point of the ungauged symmetry corresponds to a point with vanishing
Killing vector. For lorentzian signature, the surface carried into the
fixed
point by the isometry will be a null surface (the norm of the Killing
vector is conserved), in general nonsingular and hence a horizon. We see
that
fixed points of symmetry transformations generically give rise to metric
singularities when the symmetry is gauged and to horizons when ungauged.

The simplest class of coset models with a single
timelike coordinate and any number of spacelike coordinates
are the $SL(2,\IR)\otimes SO(1,1)^{D-2}\big/SO(1,1)$ models. In order to
find the metric in the large $k$ limit, we employ the standard procedure
in nonlinear $\sigma$--models \vhol,
i.e.\ find a parametrization of
the $G$ group elements, impose a unitary type gauge on the fields in the
$\sigma$--model action and then solve for the (non-propagating) $H$-gauge
fields to derive the $G/H$ worldsheet action. From that action
we can read off the corresponding background fields from \efft.
The results are that for the vector gauging, the metric is the direct
product of the  2--d black hole  of \witten\ with metric
$\d s^2=-\d a\,\d b/(1-ab)$, (see \fig\ftdbh{\tdbhtext} ) times $\IR^{D-2}$.
 That is a $D-2$
black--brane.

\figinsert\ftdbh{\epsfxsize2in$$\epsfbox{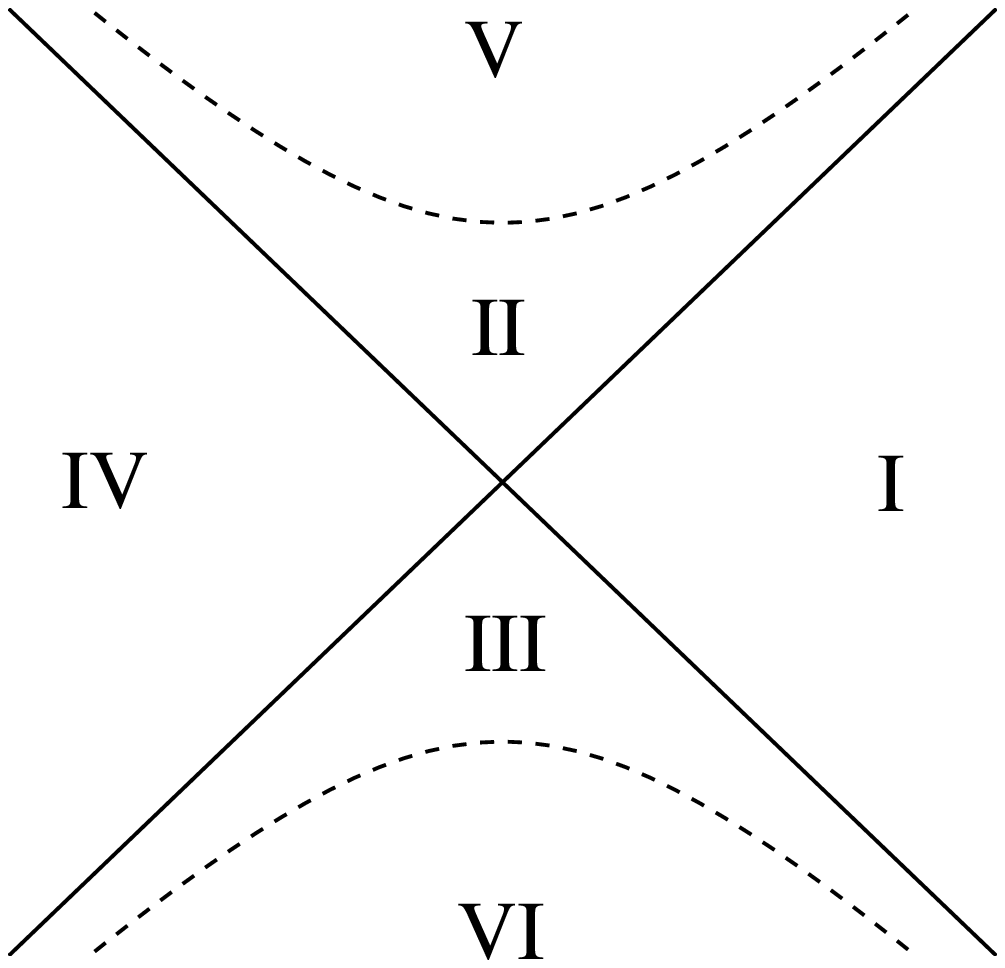}$$}{\tdbhtext}

\def\tdbstext{A two dimensional slice of the three dimensional black string
metric. In addition to the regions of \ftdbh, the regions VII,VIII
lie between the singularities and inner horizons.}
 For the axial gauging the geometry is a
3--d black string times $\IR^{D-3}$ \hh,\gique\ with two horizons and a
 singularity
as shown in \fig\ftdbs{\tdbstext}.

We should remark at this point that these solutions are only valid for a
very large $k$ limit. There has been some progress towards infering
the exact form of the metric both from the operator approach
 and from an exact treatment of the gauged WZW models. We refer the reader
 to \kt\ for a discussion of these methods.

%\figinsert\ftdbs{\epsfxsize2in$$\epsfbox{2dbs.ps}$$}{\tdbstext}

\newsec{ Abelian Duality}

We will briefly review here the standard duality corresponding
to backgrounds with abelian
isometries.
 The worldsheet action for the bosonic string in a
background with $N$ commuting isometries, can be written as
\eqn\sigmad{\eqalign{S = {1\over{4\pi\alpha'}}\int \d^2z\,
\Bigl(Q_{\mu\nu}(X_{\alpha})\,\del X^{\mu} \delbar X^{\nu}
 + Q_{\mu n}(X_{\alpha})\del X^{\mu} \delbar
X^n  + Q_{n \mu}(X_\alpha)\del X^n \delbar X^{\mu}
 \cr \qquad + Q_{mn}(X_\alpha)\del X^m \delbar X^n +
  {\alpha'\over 2} R^{(2)}\Phi(X_\alpha)\Bigr)\ ,\cr}}
where $Q_{MN}\equiv G_{MN}+B_{MN}$ and lower case
latin indices $m,n$ label the isometry directions. Since the
action \sigmad\ depends on the $X^m$
only through their derivatives, we can write it in
first order form by introducing
variables $A^m$ and adding an extra term to the action
\hbox{$\L_m(\del \overline A^m - \delbar A^m)$} which imposes
the constraint $A^m=\del X^m$.
Integrating over the Lagrange multipliers $\L_m$ returns
us to the original action
\sigmad. On the other hand
performing partial integration and solving for
$A^m$ and $\overline A^m$, we find the dual action $S'$
which has an identical form to $S$ but with
the dual background given by \refs{\busc, \gique}
\eqn\qp{\eqalign{Q'_{mn}& = (Q\inv)_{mn}\cr
  Q'_{\mu \nu}& = Q_{\mu \nu} - Q_{\mu m}\,(Q\inv)^{mn}\,Q_{n\nu}\cr
  Q'_{n \mu}& = (Q\inv)_n^{\ m}\, Q_{m \mu}\cr
  Q'_{\mu n}& = - Q_{\mu m}\,(Q\inv)^m_{\ n}\ .\cr}}
To preserve conformal invariance, it can be seen  \refs{\busc,
\giva} that the dilaton field has to transform as
$\Phi' = \Phi - \log\det G_{mn}$.
Equations \qp\ reduce to the usual
duality transformations for the toroidal compactifications
 in the case $Q_{m\mu}=Q_{\mu m}=0$ and can map
 a space with no
torsion ($Q_{m\mu} =Q_{\mu m}$)  to a space
with torsion ($Q'_{m\mu}=-Q'_{\mu m}$).

\figinsert\ftdbs{\epsfxsize2in$$\epsfbox{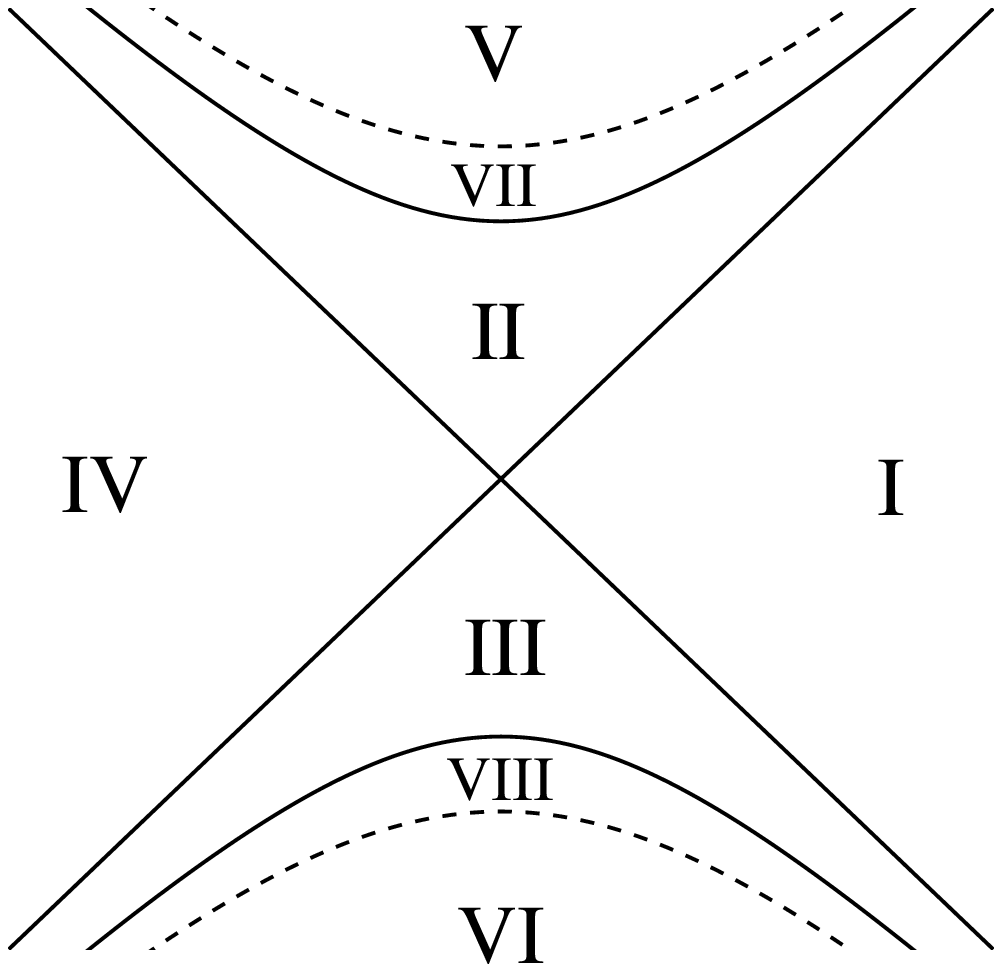}$$}{\tdbstext}

An equivalent interpretation of the duality process
just described is given by gauging the symmetry,
$\del X^m$ with $D X^m = \del X^m + A^m$ and the
term $\int d^2z\ \L_m(\del \overline A^m - \delbar A^m)$
is added to the action.  This extra term
imposes the vanishing of the field strength $F$
of the gauge fields after integration over the
Lagrange multipliers $\L_m$.
This implies that locally the gauge field must be pure gauge,
$A^m = \del X^m$.
The gauge fixing can be done either by
choosing the gauge fields to vanish or by taking
$X^m=0$ (a unitary gauge). In both cases this
reproduces the original action.
The dual theory is obtained by instead integrating out the
gauge fields and then fixing the gauge.

In \kir\ it was shown that the axial and vector gaugings of WZW
models are related by transformations like \qp\ and therefore
lead to dual geometries.
In order to make a connection between duality in this formulation and
the vector--axial duality
in $G/H$ WZW models, we just have to identify the correct action
of the vector (axial) isometry when gauging the axial (vector)
transformation
and apply \qp. It is straightforward
to see that for the $SL(2,\IR)\otimes SO(1,1)^{D-2}\big/SO(1,1)$
models for which we see that
regions
II  of both geometries are interchanged. Region
V of \ftdbh\ is mapped to region I of \ftdbs\
and in particular the singularity of the first is mapped to one
horizon
in the second. Also, region I of the vector gauging black hole gets
mapped
to regions V and VII together of the axial gauging black hole.
This has
the interesting implication that a surface which in one geometry
is perfectly
regular ($ab=\rho^2$) is mapped to the singularity in the
other geometry ($uv=1+\rho^2$).  In this case it can be seen
 explicitly that string
theory can deal with spacetimes that have singularities at
the classical level,
in the sense that there still exists a description of
interactions, etc.\ for
that region of spacetime by going to the dual geometry!

It is then interesting to investigate if these properties hold for
physically more interesting objects such as 4--d black holes.
 In
particular the Schwarzschild
metric
\eqn\schwars{ds^2=-(1-{2M/r}) dt^2 + {dr^2\over 1-{2M/r}}
                    + r^2(d\theta^2 + \sin^2\theta d\phi^2)\ ,}
times any CFT with $c= 22$ is a solution of \einst--\dil.
Direct application of the standard duality
transformation to \schwars\ for time
translations, gives the dual metric
\eqn\schwarsdu{ds^2=-{dt^2\over 1-{2M/r}} + {dr^2\over 1-{2M/r}}
                 + r^2(d\theta^2 + \sin^2\theta d\phi^2)\ ,}
with the dilaton field now given by $\Phi'=\Phi - \log(1-{2M/r})$.
This metric defines a geometry with naked singularities at
$r=0$ and $r=2M$, as it can be verified by
computing the curvature scalar
$R={4M^2\over (2M-r)r^3}$. It is easy to check
that equations	\einst--\dil\  are satisfied by
the dual metric and dilaton $\Phi'$. We have then
found a spherically symmetric solution of the
string background equations, which is not a black hole, but has naked
singularities  and
is dual to the Schwarzschild solution.

A similar analysis can be done for the 4D charged
dilatonic black holes of reference \gibbons. In this case the metric is
\eqn\cdbh{ds^2=
     -{(1-{2M/r})\over (1-Q^2/Mr)} dt^2 + {dr^2\over (1-{2M/r})
                   (1-Q^2/Mr)}
                    + r^2(d\theta^2 + \sin^2\theta d\phi^2)\ ,}
the dilaton field is $\Phi = -\log (1-Q^2/Mr)$ and the electric field
$F_{tr}= e^{\Phi}Q/(2r^2)$. It is very interesting
to note that the dual of this
solution with respect to time translations gives exactly the
same solution except that the mass parameter $M$ changes into
$ Q^2/2M$, therefore it relates
the black hole domain $Q^2<2M^2$ to
the naked singularity domain $Q^2>2M^2$.
In particular, the extremal solution $Q^2=2M^2$
is selfdual.
Notice however that the isometry group of  these 4--d geometries
is given by time translations  together
with the $SO(3)$ space rotations,
which is not abelian. It is then natural to  inquire if there is a duality
transformation associated to the existence of non--abelian isometries.

\newsec{Non--Abelian Duality }

Consider  the $\sigma$--model action \sigmad\ and
assume that the target space metric has a group $\cal{G}$ of
non--abelian isometries.  In this case, $Q_{MN}$ {\it does}
depend on $X^m$ and transforms accordingly
under $X^m\to g^m_{\ n} X^n\ , g\in \cal{G}$.
  We gauge the symmetry corresponding to a
subgroup $H\subseteq\cal{G}$
    $\del X^m \to DX^m = \del X^m +
              A^{\alpha}(T_{\alpha})^m_{\ n} X^n\ $,
and add to the action the term
$\int d^2 z\ tr(\L F) =
                        \int d^2 z\ \L_{\alpha}F^{\alpha} ,$
where in this case the gauge field strength is,
in matrix notation,
$F =
       \del \overline A - \delbar A + [A,\overline A] $.
The  $N \times N$ matrices $T_{\alpha}$ form an
adjoint representation of the group $H$.
In the path integral we have then
\eqn\pathint{
   %{\cal P} &= \int {{\cal D}X\over V_{\cal G}}\  e^{- iS[X]}\cr
    = \int {{\cal D}X\over V_{\cal G}}\
         \int D\L\ DA\ D\overline A\
       \exp\left\{ -i\left( S_{gauged}[X,A,\overline A] +
                 \int d^2 z\ tr(\L F) \right) \right\} \ ,}
where $V_{\cal G}$ is the ``volume'' of the group of isometries
and ${\cal D}X$ is the measure that gives the correct volume
element
${\cal D}X = DX \sqrt{G} e^{-\Phi}\ $.
Similar to the abelian case, the original action is obtained by
integrating out the Lagrange multiplier $\L$.  Locally, this forces
the gauge field to be pure gauge
$A = h^{-1}\del h\ ,
    \ \overline A = h^{-1}\delbar h,\ h\in H $.
By fixing the gauge with the choice
$A = 0$, $\overline A = 0$ we reproduce
the original theory.
The dual theory is obtained by integrating out
the gauge fields in the path integral \pathint.
Integrating over the gauge fields $A$, $\overline A$ we obtain
\eqn\ppredualPI{{\cal P} =
               \int {\cal D}X\ D\L\
	 \delta[{\cal F}]\ \det{{\delta{\cal F}}\over{\delta\omega}}\
                              e^{-iS'[X,\L]} \det (f^{-1})\ ,}
 where ${\cal F}$ is the gauge fixing function and $\omega$ are the
parameters of the group of isometries.  The matrix $f$ is the coefficient
 of the quadratic term in the
gauge fields \deque\ and $S'$ is given by
\eqn\predualS{S'[X,\L] = S[X]
               - {1\over{4\pi\alpha'}}\int d^2 z\
            \overline h_{\alpha} (f^{-1})^{\alpha\beta} h_{\beta}\ .}
Here  $h$ and $\overline h$
are the currents coupled to $\overline A$ and $A$ respectively \deque.
After the gauge fixing,
denoting the new coordinates in the dual manifold collectively by $Y$
we have
\eqn\dualPI{{\cal P} =
        \int {\cal D}Y\ e^{-iS'[Y]} \det (f(Y)^{-1}) \ .}
The Fadeev--Popov determinant
in the path integral contributes to the measure such that
the correct volume element for the dual manifold is
obtained
${\cal D}Y = DY \sqrt{G'} e^{-\Phi'}\ $.
The factor $\det (f^{-1})$ in the partition function
can be computed using standard
regularization techniques \busc. It generates
a new local term in the action of the form
${1\over{8\pi\alpha'}}\int d^2 z\
         \alpha' R^{(2)}\ (\Delta\Phi ) $ ,
which corresponds to the change in the dilaton
due to the duality transformation
\eqn\newdil{\Phi' = \Phi - \log\det f\ .}
This change in the dilaton transformation is the
shift necessary to retain the conformal invariance
of the dual theory. The requirement that the correct volume element
is obtained in the dual theory means that
$e^{-\Phi'}= \left[ e^{-\Phi}\ \sqrt{G\over {G'}}\
             \det {{\delta {\cal F}}\over {\delta\omega}}
	        \right]_{{\cal F}=0} $
This prescription reduces to  the one
for abelian duality because  in that case
the Fadeev--Popov determinant is trivial. A consistency
check of the change in the dilaton is obtained by comparing both
expressions. Notice that our assumption that the action of the group is
linear on the coordinates was made only for simplicity and it is not
necessary. We  can actually rederive
the duality transformations in a coordinate independent way, as it was
done in \hs\ for any gauged sigma model.

In general, we cannot write explicitly the gauge fixed
dual action. Therefore, we are not able to
present the new metric and antisymmetric tensor fields in a closed
form, as was done for the abelian case in equations \qp.
As an example, let us
consider a theory  for which the target space metric
has a maximally symmetric subspace
with ${\cal G} = SO(3)$ and no
antisymmetric tensor. The coordinates $X^M, M = 1,...,D$,
can be decomposed into the two angular coordinates
($\theta$, $\phi$) describing
$2$--dimensional spheres, and $D-2$ extra
coordinates ($v^\mu$) specifying the different spheres in
the $D$ dimensional spacetime. The action takes
 the form
\eqn\sots{S[v,\theta,\varphi] = S[v] + \int d^2 z\ a^2\Omega(v)
  \left(\del\theta \delbar\theta +
          \sin^2\theta \del\varphi\delbar\varphi
                        \right)\ .}
It is simpler to
treat the coordinates $\theta$  and $\phi$ in terms
of cartesian coordinates $X^m$ in
3--dimensional space on which $SO(3)$ can
act linearly, so
we write the $\s$ model action in the form
\eqn\sonS{\eqalign{S[v,X] &= S[v]+\int d^2 z\ \Omega(v)
      \left\{ g_{mn}\del X^m \delbar X^n +
  {1\over{2a\sqrt{\Omega}}}\lambda (g_{mn} X^m X^n - a^2)\right\}\cr
     &+ {1\over{8\pi\alpha'}}\int d^2 z\
         \alpha' R^{(2)}\ \Phi\ ,\cr}}
where
$S[v] = \int d^2 z g_{\mu\nu}(v)\del v^{\mu} \delbar v^{\nu}$,
the metric $g_{mn}$ is
diagonal and constant and
the Lagrange multiplier term fixes
the 3 dimensional space to be
a sphere of radius $a$.
Gauging this action and fixing the gauge $A=\bar A=0$ we obtain
the original action.

A convenient choice of gauge is to set
$X^1=X^2 = 0, \quad X^3 = a$, $A^1 = \del\theta$,
$A^2 = - \sin\theta\del\varphi$ and
$A^3 = \cos\theta\del\varphi$) leading to the original action in
spherical coordinates \sots.
%%and then use the remaining $SO(N-1)$ gauge freedom to gauge away ${1\over2}
%%%(N - 1) (N - 2)$ of the Lagrange
%multipliers $\L_{\alpha}$.  We then obtain
%
We can write a general expression for the dual action after fixing the
coordinates
$X^m$ as above, but before fixing the remaining degree of freedom
corresponding to one of the Lagrange multipliers $\Lambda_\alpha$.
We then have
\eqn\dualsons{S^{dual}[v,\L] = S[v]+{1\over{4\pi\alpha'}}
                 \int d^2 z\ \left(
               \del\L_{\alpha} (f^{-1})^{\alpha\beta} \delbar\L_{\beta}
                   \right) +
          {1\over{8\pi}}\int d^2 z\ R^{(2)}\Phi' \ .}
{}From this expression,  we can in principle read off
the new background fields
as in \qp. This is actually the general expression for any group.
 But we  still  have to complete
the gauge fixing for the $\L_{\alpha}$. In our case,
 Choosing $\L_2 = 0$ and defining
$ x^2 = \L_1^{\ 2} + \L_3^{\ 2}$ and $y = \L_3$, we obtain
the dual theory action
\eqn\dualsothrees{\eqalign{
    S^{dual}[v,x,y] = & S[v]+{1\over {4\pi\alpha '}} \int d^2 z\
          {1\over {a^2\Omega(v) (x^2 - y^2)}}\
    \left( a^4 \Omega(v)^2\del y \delbar y
                    + x^2 \del x \delbar x \right) \cr
  &\qquad + {1\over {8\pi}} \int d^2 z\ R^{(2)}\Phi'\ ,\cr}}
where
$\Phi ' = \Phi - \log [a^2 \Omega(v)\ (x^2 - y^2)] $.

We will now present,  some 4D black hole
backgrounds
and their duals. Consider the dual geometry of \schwars\ with
respect to the
$SO(3)$ symmetry.  We find
\eqn\schwarsdd{ds^2=-(1-{2M/r}) dt^2 + {dr^2\over 1-{2M/r}}
   +{1\over {r^2 (x^2 - y^2)}}\ [r^4 dy^2
                    + x^2 dx^2]\ ,}
with the new dilaton $\Phi' = \Phi - \log [r^2 (x^2 - y^2)]$. The
regions $x=y$
and $r=0$ are real singularities whereas
$r=2M$ is only a metric singularity
corresponding to a horizon as in the original case.
Notice that the metric \schwarsdd\ is {\it not}
spherically symmetric,
in fact its only isometry is time translations. Neither is it
 asymptotically
flat. For large $r$, the $x$ dimension gets
squeezed and the other dimensions
behave like a $2+1$ dimensional space--time.
The surfaces $x = {\rm constant}$ are just
$2+1$ dimensional black holes away from the singularities
$\sin\theta=0$, $(y=\pm x).$
Again, it is straightforward to check that this solution
satisfies equations \einst--\dil\
thus providing new string vacua.
To find new solutions, we can certainly combine
both dualities above.  We can also consider different
coordinate systems. For instance, using the Eddington-Finkelstein
instead of the Schwarzschild coordinate system,
%we can
%find its dual with respect to time translations.
the dual metric with respect to time translations
is identical to \schwarsdu, but  there is also
  torsion illustrating that duality does not commute with
coordinate transformations.

\newsec{Conclusions}

We have shown some of the interesting properties that string theory
backgrounds
have due to the existence of duality symmetries.
In particular we have presented a general non--abelian duality symmetry
which reduces to the standard one when the gauged group is abelian
and allows to find new string backgrounds at least to lowest order
in $\alpha'$. It is not clear if this symmetry will survive
beyond string tree--level though. We have shown examples where
this symmetry generates new vacua but the list of possible
applications is obviously very large. An interesting possibility is to
study the dual geometries to Friedmann--Robertson--Walker
cosmologies. Also,
 in  10--d Minkowski
space there is
a  large group of  isometries  which we can use to
find
 the dual geometries to flat spacetime.
 Furthermore,
consider any of the `4--d strings'  with spacetime
4--d
Minkowski space, for which billions of solutions are known.
Again
the isometry group of this space allows the existence of
 many new
 vacua, duals of 4--d strings.
It would also be interesting to find
 the
moduli space of the solutions connected by this new duality,
 analogous to the
$SO(N,N)$ continuous transformations which generates the
 moduli
space for the case of abelian duality \cfg, \mv.
Let us finish with a curiosity. The process we followed for
duality can be applied to any 2--d theory. In particular we
can apply it to the free Dirac action which will have global symmetries.
By gauging them and setting the field strength to zero, we can find
the dual theory. In the abelian case, the gauge field equations yield
to $\bar\psi \gamma^\mu \psi = \epsilon^{\mu\nu}\del_\nu \L $
(where $\L$ is again the Lagrange multiplier). These are just the
relations satisfied by the conserved currents in abelian bosonization.
The explicit solution for the fermions requires the standard
bosonization technology though. Also,
for the non--abelian case the relation to the WZW model is not yet
 clear from this point of view but it is interesting to note that duality
 and bosonization in 2--d systems could be understood from
the same underlying principle. We believe that other duality symmetries
such
as mirror symmetry should also be understood from this approach.
\medskip

{\noindent\bf Acknowledgements}
\smallskip

The  work presented here is the result of
collaborations with X. de la Ossa and P.
Ginsparg. I would also like to thank the organizers for such an
enjoyable workshop. This work has been supported by the Swiss
National Science Foundation.

%\listrefs
%\listfigs   %(if necessary)
%\bye
\nobreak
\listrefs
%\ifx\answ\bigans\else\quad\vfill\eject\fi
\bye